# A FRAMEWORK FOR BUSINESS INTELLIGENCE APPLICATION USING ONTOLOGICAL CLASSIFICATION

A. Martin

Research Scholar, Department of Banking Technology, Pondicherry University, Pondicherry

D.Maladhy

Department of computer Science and Engineering, Sri Manakula Vinayagar Engineering College, Madagadipet.Pondicherry University, Pondicherry

Dr.V.Prasanna Venkatesan

Dept. of Banking Technology, Pondicherry University, Pondicherry

**Abstract:**

**Every business needs knowledge about their competitors to survive better. One of the information repositories is web. Retrieving Specific information from the web is challenging. An Ontological model is developed to capture specific information by using web semantics. From the Ontology model, the relations between the data are mined using decision tree. From all these a new framework is developed for Business Intelligence.**

1. Introduction

Nowadays, there is a growing interest in information extraction (IE). It is of paramount importance in several real world applications in the areas of business, competitive and military intelligence. The growth of the web has dramatically expanded the range and volume of information that are relevant to the organization's business. Information ranging from contracts to won new products are daily reported online news sources.

Having sufficient amount of information has become a key factor of success in all fields of human activity. The information integrated from various online news sources is required for the support of decision making process for business. So we need a BI model to capture rich semantic representation from news sources.

Many organizations monitor the online news from various news sources for information needed for their business insights. So we need an ontology model to capture online news from different news sources. This task is performed by generating phrases from these representations and matching these phrases against the news using a set of syntactic and semantic transformation. The representation that best matches is taken as relevant information need for their business.

This paper is organized as follows. Section 2 related work describes ontology concepts, business intelligence. Section 3 Information Retrieval Section 4 presents our approach using ontology in this paper. Finally, Section 5 presents conclusions.

2. Related Work

2.1. Creation of Ontology Model

Ontology consists of concepts, attributes, and properties representing relationships between concepts. Ontology's properties can represent user-defined relationships as well as is-a and has a relationships. Defining semantic relationships between concepts enables the development of a machine system that can automatically interpret and understand the meaning of concepts used in ontologies. For such automatic knowledge acquisition, anthologies have been widely studied and developed in a variety of domains, which increases the need for sharing and reusing available ontologies [4].

However, the task of sharing and reusing ontologies has many challenges. First, there is a problem related to polysemy and synonym. A word can have multiple meanings (i.e., polysemy) or there may be more than one word that means the same thing in a domain (i.e., synonyms), which causes difficulty in ontology matching. Second, polymorphism in expressions also brings  about a difficulty in finding semantically related concepts. For example, an ontology might use noun forms of names and another ontology uses adjectival forms of names to indicate the same concept. Third, ontologies do not give a restriction on the count of a concept to represent a meaning. For example, in order to represent someone's name, an ontology simply has a concept–'Name', while another ontologies use two concepts 'FirstName' and 'LastName' [5].

Finally, both a concept and a property can be used to give a meaning. For example, an ontology can use the concept 'adviser' and another can use the property 'advised By' to indicate someone who gives advice [4]. Due to such difficulties, ontology matching has been the subject of considerable research. In particular, as users increasingly demand a way to automatically exchange information and knowledge in applications, the automation of ontology matching becomes a research issue of great importance [2].

An Ontology matching is generally comprised of three phases of similarity measure – similarity measure between concepts, similarity measure between properties, and logical inference similarity measure; and there are four major dimensions of similarity, i.e., lexical similarity, structural similarity, instance similarity, and logical inference similarity. Existing ontology matching methods rely heavily on lexical similarity. Lexical similarity generally employs the character comparison method that focuses only on order of characters, irrespective of their meanings. However, this approach cannot solve the problem of polysemy and synonym described above.[4]

Ontology is knowledge representation method, it can describe the concepts, relations between the concepts and axiom in formal way, ontology provides a sound semantic ground of machine-understandable description of digital content .Ontology is very suitable to express domain knowledge [3]

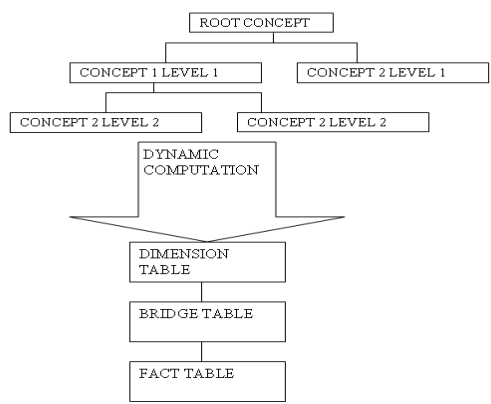

Fig. 1. The Two-Steps Indexing Process.

To preliminary indexing the information system content we adopt a semantic indexing engine. The indexing process that it implements is a two-step process of non structured documents, divided in the three layers illustrated in Figure 1. During the first step, from the Unstructured Docs layer to the Terms Set layer, the engine, based on the API of the Open Source search engine Lucene, indexed each document, thus obtaining a set of index-terms. In the second step, from the terms set layer to the ontologies layer, these terms were contextualized and associated with the concepts of predefined ontologies.

In order to effectively run our process, the following assumptions were imposed:

• The concepts included in the dictionary were exclusively linked by hypernymy and hyponymy relations;

• Each Ontology was based on a hierarchic structure.
The aforesaid restrictions obviously entailed an experimentation that was to be limited to the context, albeit it could also go for other real cases.

## 2.2. Business Intelligence

Business Intelligence (BI) solutions offer the means to transform data into information and derive knowledge through analytical tools in order to support decision making. Analytical tools should support decision makers to find the right information quickly and enable them to make well-informed decisions. Business intelligence (BI) is the process of gathering enough of the right information in the right manner at the right time, and delivering the right results to the right people for decision-making purposes so that it can continue to yield real business benefits, or have a positive impact on business strategy, tactics, and operations in the enterprises.[4]Business Intelligence (BI) is defined as an integrated set of tools to support the transformation of data into information in order to support decision making.[2]

### 2.2.1Business Intelligence Components

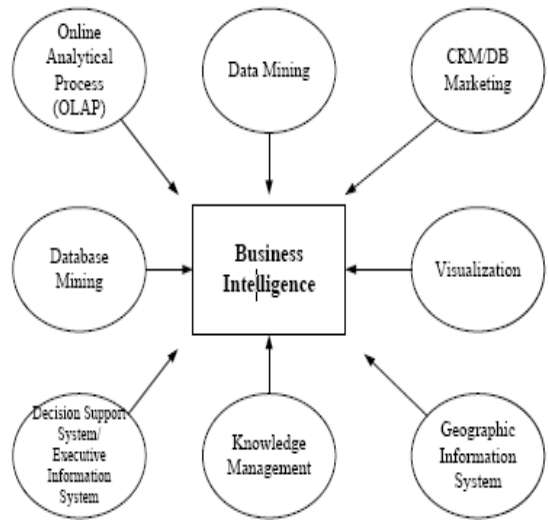

Fig. 2. BI Components

The BI components, which comprise online analytical process (OLAP), knowledge management, CRM/database marketing, database mining, visualization, decision support system/executive information system, data mining (DM) and geographic information system (GIS)

### 2.2.2 How Business Intelligence works?
The BI process, which works to transform data into information, and then knowledge using analytical tools, such as DM, OLAP, visualization etc. Finally, the generated knowledge can be used to support business decisions. [6]

A typical BI architecture contains a Data Warehouse (DW), an Extraction, Transformation and Loading tool (ETL) and a set of analytical tools. DW is an integrated repository of data consolidated from different data sources through ETL tools. Usually, the approach used for data modelling in DW is the star schema, which defines that descriptions of the business (e.g. product description) are stored in dimensions, while the measures (e.g. amount of items sold) are kept in fact tables. DW supplies the data that is presented to the user through analytical tools. Different kinds of analytical tools such as On-Line Analytical Processing (OLAP) are used to provide the means for users to define their analyses (i.e. reports or cubes) and explore the results through analytical functionalities [2].ETL (Extract-Transform-Load) can integrate and increase the value of data according to the uniformed rules. After data extract, data clearing and data arrangement, data will be loaded into the data warehouse, which is pre-defined precisely. In brief, ETL is a transfer process from data source to the target data warehouse and an important step of data warehouse implementation. From the practical experience at home and abroad, it is a common sense that ETL rules design and implementation is a key element in BI project.

The main module of ETL Accurate way to design ETL process to make it efficient, flexible and maintainability. ETL can be divided into five modules: data extraction, data validation, data cleaning, data conversion and data loading.

3. Information Retrieval
3.1. Information Retrieval by Search engines
Information Retrieval is the task of identifying documents in a collection on the basis of properties described to the documents by the user requesting the retrieval [2]. Current information retrieval techniques on web are not intelligent enough to exploit the meaning of data i.e.; semantic knowledge within documents and hence cannot give precise answers to precise questions. Information retrieval may be expressed in the form as shown in figure 3. Different types of data are available on the Web viz, structured, unstructured and semi structured. Structured data is in the form of text document. Semi structured data are not full and grammatical text. [9]

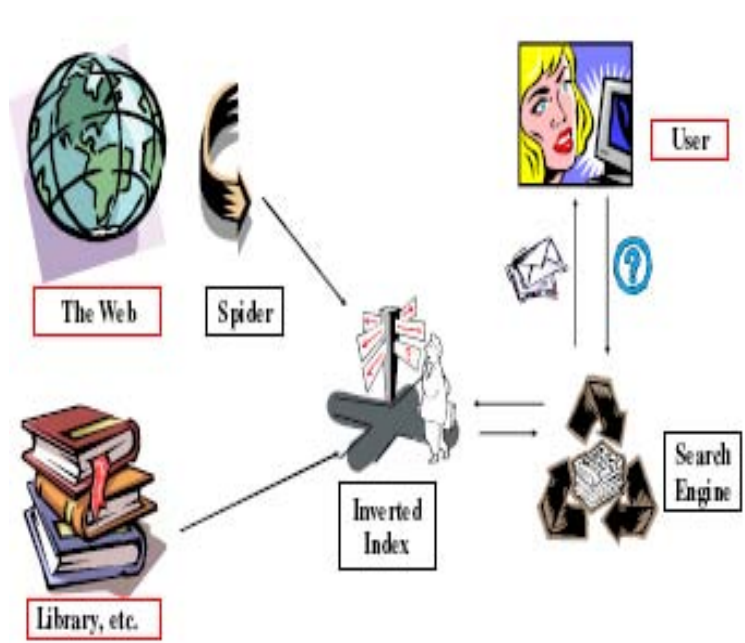

Fig. 3. Information Retrieval Method

The figure .3 shows that user gives Keywords to the search engine which in turns search the keyword in the index which in turn retrieves information from the library.
3.2. Information Retrieval using Ontology
An example of the information retrieval using the ontology shows each node represents a word that has a hub weight and a related document list. We can retrieve the document using information such as relations and document list of our ontology. As a general rule, a word which appears in many documents is not considered as being important in individual documents. Therefore, the hub words appeared in many documents are not useful for a set of representative keywords of the each documents. A hub word plays a role as a bridge linking one index term to another index term by relations.[6]

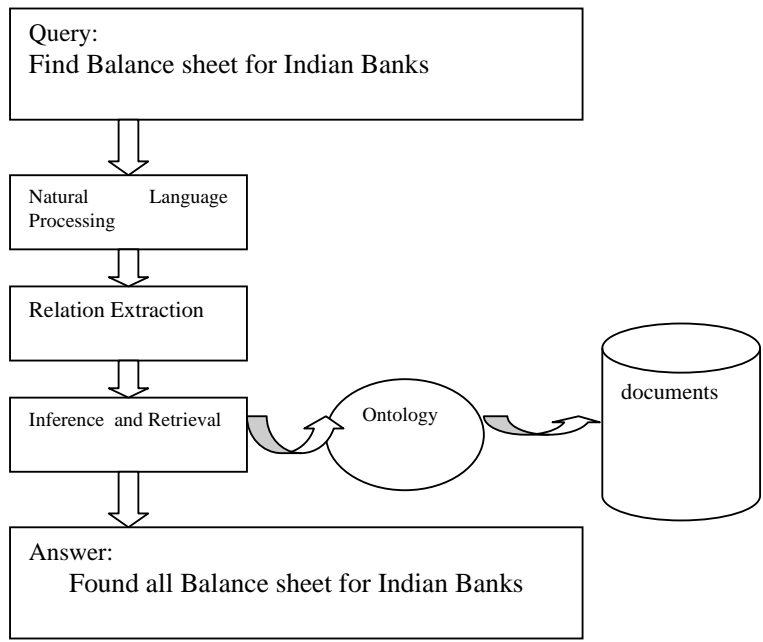

Fig.4. An example of the information retrieval using the ontology

## 4. Our Approach

Our approach captures semantic representations from news sources by generating phrases from representations of interest and matching these phrases against the news using a set of syntactic and semantic transformations. This project presents a new method towards ontology matching. This captures semantics, for searching relevant information from online news source through ontology. This is performed by generating phrase from the representation and matching the phrases against the news using concept matching algorithm. The representation that best matches is taken as relevant information need for their business.

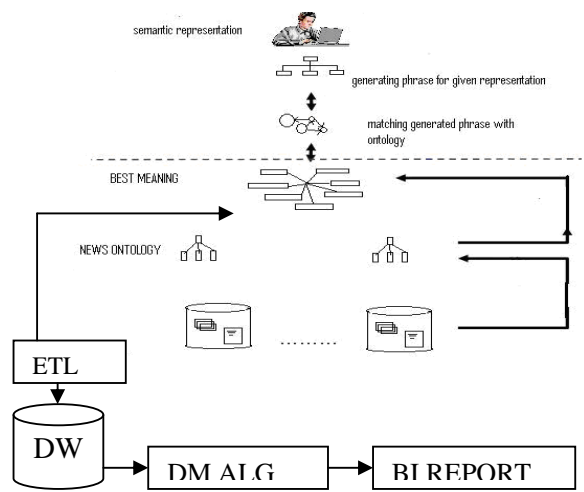

Fig. 5.Proposed System

The Figure 5 shows the proposed architecture for business intelligence application .semantic captured as explained above techniques using ontology. We crawl all information about online news for business intelligence application .This diagram shows overall architecture which consist of

- Collection of best meanings gathered from ontology
- ETL
- Database
- Data mining Algorithm
- Front End

## 4.1 Collection of best meanings gathered from ontology

The keywords  taken as input for example: Find the information about education loan available in all banks will tokenized as Find,information,,eduction ,loan, bank and phrase will be generated by lexical replacement with the component library which contains framenet,wordnet. The generated phrase can matched with ontology by using following matching algorithm.

## 4.2 Matching algorithm

Definition 1 (weight vector): An Ontology is made up of multiple terms (which are called concepts) that are related and constrained by various structural frameworks. Ontology of n concepts is mapped into the vector ( $r_1$ ,$r_2$ ,...,$r_n$ ) by matching rule $r_i \in [0,1]$ . The value of $r_i$ denotes the influence of the ith concept on whole Ontology semantic and is decided by matching rule. The vector ($r1$, $r2$,...,$r_n$ ) is called weight vector.

Definition 2 (concepts--weights vectors): Concepts
set ($C_1$ ,$C_2$ ,...,$C_n$ ), where $C_i$  is the ith concept of the Ontology, and corresponding weight vector ( $r_1$ ,$r_2$,...,$r_n$ )  are named concepts--- weights vectors together.

Definition 3 (Input): Generated phrase

Definition 4 (evaluating Ontology): An Ontology that needs to compare with input..

Matching algorithm need to decide weight vector according to matching rule then map concepts set to obtain result vector. The basic idea of the algorithm is to estimate semantic similarity between the input message and preliminary keywords based results (They are regarded as input  and evaluating Ontologies respectively). The algorithm is the following.

Algorithm

Input: Input, evaluating Ontology

Output: result vector ( $R_1$ ,$R_2$ ,...,$R_n$ ) parse  Input  and Ontology  into ( $I_1$ ,$I_2$ ,...,$I_m$ ) and ($C_1$ ,$C_2$ ,...,$C_n$ )  ;
create corresponding weight vectors ( $t_1$ ,$t_2$ ,...,$t_m$ )  and ( $r_1$ ,$r_2$ ,...,$r_n$ )  by matching rules;
for all $I_i$ in ( $I_1$ ,$I_2$ ,...,$I_m$ )  do  compare  $I_i$ with $C_j$ in ($C_1$ ,$C_2$ ,...,$C_n$ );
if  $I_i = C_j$ then $R_i = t_i \times r_j$ ;
else  $R_i = 0$ ;
end if
end for
output result vector ( $R_1$ ,$R_2$ ,...,$R_m$ ).[8]

## 4.3 Decision Tree

A decision tree is a classification scheme which generates a tree and a set of rules, representing the model of different classes, from a given data set. The set of records available for developing classification methods is generally divided into two disjoint subsets-a training set and a test set [1].

The C4.5 system consists of four principal programs:
1) Decision tree generator ('c4.5') - construct the decision tree
2) Production rule generator ('c4.5rules') - form production rules from unpruned tree
3) Decision tree interpreter ('consult') - classify items using a decision tree
4) Production rule interpreter ('consultr') - classify items using a rule set

C4.5 algorithm performs well in constructing decision trees and extracting rules from the bank news C4.5 algorithm is needed in order to provide ease of use and better visualization of the decision trees for the users. In order to discover new, meaningful and "actionable" knowledge from the banking news, more data needs to be collected. Our approach after storing the bank news result  in data warehouse ,the mining process will started  finding the steps needed to utilize C4.5 Algorithm for classifying  banking dataset, discovering rules generated from the dataset and the meaning of them.The Steps of C4.5 algorithm
Steps of algorithm steps:

1)Start from the Root node, set the current node C is the root node, at this point all objects belong to the collection of objects C.

2)If all of the C's objects belong to the same type, set node C is this category, and then stop ,otherwise continue do step 3.

3)Or else, if the records is empty, then return to leaf node whose status is none loss.

4)Otherwise, select the smallest entropy properties, set A(or directly specify that attributes),as the root node, for all yet to emerge from the root to the current nodes the path attribute Ai (called candidate attributes),divide up C based on Ai, and calculate E (Ai) and information profit.

5)For each possible value of attribute, set V;set C is the sample subset that A=Vi; increase the branch from the root node(A=Vi),now, the data become C1,attributes become records-(A);when $pj < 0.1\%$,pj come from C1;

6)If the attribute is empty, then create leaf nodes whose label is non-loss.

7)Otherwise, create a recursively sub-tree ICDT(Ci，records-{A})；

8)End (All values of A);

9)End of the recursive function.

### 4.4 Discussion

The discussion between the original model and our approach shows in the below table 1.The table represents the difference between Google search and our search. The query: Balance sheet for Indian Banks is placed and the results shown in Existing are 2,250,000 and the duration is 0.10.In our approach the result will be relevant and it gives less than the results found in Existing.
The evaluation between the original model and our approach shows in the below table 1

Table.1 Page retrieval result in inverted index

| S.No. | Query searching | Original model (inverted index) | Duration (Seconds) |
|---|---|---|---|
| 1 | Balance sheet for Indian Banks. | 2,250,000 | 0.10 |
| 2 | Impact of recession in Indian banks. | 6,070,000 | 0.16 |
| 3 | Phishing and Online banking. | 1,750,000 | 0.12 |
| 4 | Ontological model for banking applications | 54,100 | 0.23 |
| 5 | Current economic status of  India | 3,080,000 | 0.11 |

The evaluation of the original model shows in the table that result found in the search engine gives the result to the maximum of 6,070,000 so it takes time to examine all the result where as our approach used with ontology tree index shows the result of relevant semantic based information with exact and minimum result.

## 5. Conclusions

Thus the given keyword will be given as input for searching relevant information from online news source through ontology. From the given input, the matching process will take place with the news by using matching algorithm with ontology. Then the best matched news will be stored in the data warehouse .By using C4.5 mining algorithm, the data is analyzed and the knowledge is obtained. Then the decision making process will take place and the BI report is generated.

**Authors**

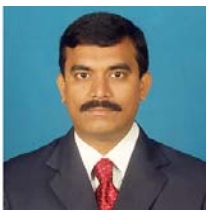
Mr.A.Martin Assistant Professor in the Department of Information Technology in Sri Manakula Vinayagar Engineering College. Pondicherry University, Pudhucherry, India. He holds a M.E and pursuing his Ph.D in Banking Technology from Pondicherry University, India. He can be reached via jayamartin@yahoo.com

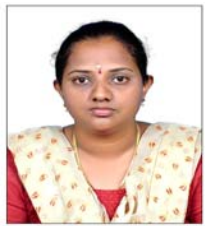
Mrs.D.Maladhy II year M.tech student in the project phase in the Department of computer Science and Engineering in Sri Manakula Vinayagar Engineering College. Pondicherry University, Pudhucherry.she holds a B.Tech can be reached via maladhymtech @gmail.com

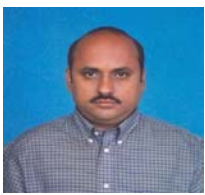
Dr.V.Prasanna Venkatesan, Associate Professor, Dept. of Banking Technology, Pondicherry University, Pondicherry. He has more than 20 years Teaching and Research experience in the field of Computer Science and Engineering; He has developed a compiler for multilingual languages, he can be reached via prasanna_v@yahoo.com.